\documentclass[twocolumn,showpacs,showkeys,
  preprintnumbers,amsmath,amssymb,page]{revtex4}
\usepackage{graphicx}
\usepackage{dcolumn}
\usepackage{bm}

\begin{document}

\preprint{}

\title{Suppression of Ecological Competition by Apex Predator}

\author{Taksu Cheon}
\email[E-mail:]{taksu.cheon@kochi-tech.ac.jp}
\author{Shigemi Ohta}
\email[E-mail:]{shigemi.ohta@kek.jp}
\affiliation{
${}^*$
Laboratory of Physics,
Kochi University of Technology,
Tosa Yamada, Kochi 782-8502, Japan
\\
${}^\dagger$
Institute of Particle and Nuclear Studies, KEK,
Tsukuba, Ibaraki 305-0801, Japan
}

\date{August 11, 2003}

\begin{abstract}
In the framework of Lotka-Volterra dynamics with evolutionary parameter variation,
it is shown that a system of two competing species which 
is evolutionarily unstable, 
if left to themselves,
is stabilized by a commmon predator preying on both of them.
Game-theoretic implications of the results are also discussed.
\end{abstract}

\pacs{87.23.-n, 05.45.-a, 89.75.Da}
\keywords{Population dynamics, Lotka-Volterra system, Game theory,
Ecological stability}
\maketitle

%
%
From the dominant plants in forest vegetation 
to the wild beasts in savanna,
an often encountered ecological paradox exists in the form of stabilizing 
influence of the top predator.
While two species in direct competition rarely form 
a stable ecosystem, they often coexist under the
dominance of a common predator \cite{DU01}.  
The most illustrative example is found in the {\em trophic pyramids},
where the apex predator,
the most savage aggressor of all, seem to act as 
the key guarantor of the stability of the whole system \cite{EL27}.
For species with intellectual capacity, this fact might be explained 
as a result of conscious effort of enlightened self-interest.
But the peacekeeping function of the apex predator is so
prevalent throughout ecosystems,
that the existence of a simple and universal dynamics should 
be suspected.

The purpose of this paper is to understand the structure and
stability of ecosystems composed of competing species
in the framework of 
evolutionary population dynamics \cite{MA74,HS88}. 
The tool we employ is the Lotka-Volterra equation
with adiabatic parameter variation \cite{CH03}. 
In this approach, the ecological dynamics is determined by the
time variation of the variables representing the population of the species,
while the adiabatic parameter variation represents 
the behavioral evolution of the species.
The viability of a species in this framework is judged both by
the short-time ecological stability of the orbit and also by the long-term 
evolutionary stability of  the shifting parameters. 

We focus specifically on a system that consists of two self-sustaining  but 
competing species and an apex predator who preys over both competitors.
We show that the system evolves towards an
evolutionarily stable configuration in which
the warring preys are tamed into the peaceful coexistence. 
This is in contrast to the case of two competitors left to themselves, 
in which there are no evolutionarily stable solution for coexistence, 
and  ``arms race''  drives one of the competitors into eventual extinction.
We also show that our results can be interpreted in a game-theoretic language
as the apex predator turning the prisoner's dilemma between two 
competitors into  a collaborative game.
%

%
%
\begin{figure}
\includegraphics[width=3.3cm]{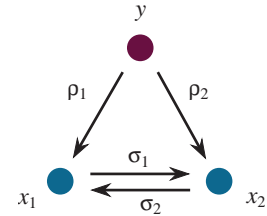}
\caption
{\label{fig1}
A symbolic diagram showing the pecking order of the three species
system described by (1).  The arrows represent the aggression and predation
with the specified intensities.
}
\end{figure}
%
%
%
Suppose there are two populations of
competing species $x_1$ and $x_2$  which are preyed upon 
by a common predator population $y$ 
(Fig.\ \ref{fig1}).
We describe the ecological dynamics of
this system by the standard Lotka-Volterra equation
%
\begin{eqnarray}
{\dot x_1} &=& b_1 x_1  - a_1 x_1^2 - \sigma_2 x_1 x_2 - \rho_1 x_1 y ,
\\ \nonumber
{\dot x_2} &=& b_2 x_2  - a_2 x_2^2 - \sigma_1 x_1 x_2 - \rho_2 x_2 y ,
\\ \nonumber
{\dot y} &=& -d y  + f \rho_1 x_1 y + f \rho_2 x_2 y .
\end{eqnarray}
Here, $b_1$, $b_2$ are the reproduction rates for species $x_1$, $x_2$,
and $a_1$, $a_2$, the environmental limitation factor to their growth. 
The coefficient $d$ is the decay rate for the predator $y$, and $f$, the efficiency of
its predation.  In the last equation, a term proportional to $y^2$ could be added
for consistency with other equations, but this can be shown to introduce simply
a technical complication without affecting the main line of our arguments.
Also, specifying separate predation efficiencies 
for $x_1$ and $x_2$ makes no essential difference, because 
the result can be turned into the original form (1)  with rescaling of variables.
The parameters $\rho_1$ and $\rho_2$ are  the aggression intensities of 
the apex predator $y$ towards $x_1$ and $x_2$, respectively.
Similarly, $\sigma_1$ and $\sigma_2$ are the
aggression intensities of $x_1$ to $x_2$ and of $x_2$ to $x_1$, respectively.
We assume all parameters to be positive real numbers.
At this stage, we treat all of them as fixed numbers, making no
distinction between the Roman denominated ``environmental'' parameters
and Greek denominated ``behavioral'' parameters.
A nontrivial fixed point $x_i(t) = X_i$, $y(t)= Y$ ($\dot X_i = \dot Y = 0$) 
with $i=1,2$
is given by
%
\begin{eqnarray}
X_1 &=&  
{1 \over f} \cdot
{  {  d  (a_2 \rho_1 - \sigma_2 \rho_2) - f (b_2 \rho_1-b_1 \rho_2)\rho_2 } 
   \over
   { a_2 \rho_1^2 + a_1 \rho_2^2 - (\sigma_1 + \sigma_2) \rho_1 \rho_2}  } ,
\\ \nonumber
X_2 &=&  
{1 \over f} \cdot
{  {   d (a_1 \rho_2 - \sigma_1 \rho_1) + f (b_2 \rho_1-b_1 \rho_2)\rho_1 } 
   \over
   { a_2 \rho_1^2 + a_1 \rho_2^2 - (\sigma_1 + \sigma_2) \rho_1 \rho_2} } ,
\\ \nonumber
Y &=&
-{d \over f} \cdot
{  {   a_1a_2 - \sigma_1 \sigma_2  } 
   \over
   { a_2 \rho_1^2 + a_1 \rho_2^2 - (\sigma_1 + \sigma_2) \rho_1 \rho_2} }
\\ \nonumber
& &
+{  {   (a_2 b_1 - b_2 \sigma_2 )\rho_1 + ( a_1 b_2 -b_1 \sigma_1)\rho_2 } 
     \over
     { a_2 \rho_1^2 + a_1 \rho_2^2 - (\sigma_1 + \sigma_2) \rho_1 \rho_2} } .
\end{eqnarray}
The stability of the fixed point is determined by the behavior of the linearized map
%
\begin{eqnarray}
M =
\begin{pmatrix}
-a_1 X_1  & -\sigma_2 X_1  & -\rho_1 X_1 \\
-\sigma_1 X_2  & -a_2 X_2  & -\rho_2 X_2 \\
 f \rho_1 Y  &  f \rho_2 Y  & 0 
\end{pmatrix} .
\end{eqnarray}
Namely, the fixed point is stable when real part of all the eigenvalues $\lambda$ of $M$
determined by
\begin{eqnarray}
\left | {\lambda I - M }\right | = 0 
\end{eqnarray}
is negative.

When the fixed point is of stable, attracting sort, neighboring orbits
form an absorbing spiral in phase space.
We now assume that evolutionary pressure of selection and adaptation 
are at work.
We can then regard aggression intensities $\rho_1$, $\rho_2$, $\sigma_1$ 
and $\sigma_2$ as {\em evolutionarily adjustable parameters}
which evolve along the path
that simultaneously increase the functions
 $X_1[\sigma_1]$, $X_2[\sigma_2]$ and $Y[\rho_1,\rho_2]$ until
they reach the optimal values.
There are several indirect pieces of evidence supporting the existence of 
this type of adiabatic evolution
among real-life ecosystems \cite{LB01, FH01}.
It is convenient to start with the maximization condition for the apex predator
 $\partial Y/ \partial\rho_1 |_{\rho_1^\star} = 0$ 
 and  $\partial Y/ \partial\rho_2 |_{\rho_2^\star} = 0$ .  We then have the
 relations
%
\begin{eqnarray}
\rho_1^\star = { d \over f  }\cdot
  { {2 a_1 a_2 b_1 - b_1 \sigma_+ \sigma_1 + a_1 b_2 \sigma_-} 
    \over 
    {a_2 b_1^2 + a_1 b_2^2 - b_1 b_2 \sigma_+} } ,
\\ \nonumber
\rho_2^\star = { d \over f  }\cdot
  { {2 a_1 a_2 b_2 - b_2 \sigma_+ \sigma_2 - a_2 b_1 \sigma_-} 
    \over 
    {a_2 b_1^2 + a_1 b_2^2 - b_1 b_2 \sigma_+} } .
\end{eqnarray}
These conditions give the expressions
%
\begin{eqnarray}
X_1^\star &=&  
    { {2 a_2 b_1- b_2 \sigma_+} \over { 4 a_1 a_2 - \sigma_+^2 } } ,
\\ \nonumber
X_2^\star &=&  
    { {2 a_1 b_2- b_1 \sigma_+} \over { 4 a_1 a_2 -\sigma_+^2 } } ,
\\ \nonumber
Y^\star &=& { f \over d } \cdot
   { {a_2 b_1^2+a_1 b_2^2- b_1b_2\sigma_+} 
     \over { 4 a_1 a_2 - \sigma_+^2 }} .
\end{eqnarray}
The quantities $-X_1^\star$ and $-X_2^\star$ 
as functions of $\sigma_1$ and $\sigma_2$
act as  the ``potential surface'' for the variation of $\sigma_1$ and $\sigma_2$.
In (5) and (6), the notation $\sigma_\pm \equiv \sigma_1\pm\sigma_2$ is used.
With the definitions $\alpha$ $\equiv \sqrt{a_1 a_2}$ 
and $\beta$ $\equiv \sqrt{a_2/a_1}\cdot b_1/ b_2$,
valid parameter range for $X_i$ and $Y$ being positive and 
stable $(\Re \lambda  < 0)$ is given by 
%
\begin{eqnarray}
\sigma_1+\sigma_2 < \min{\{\alpha\beta, \alpha/\beta\}} .
\end{eqnarray}
That stability requirement is satisfied can be checked
by the fact that all the coefficients of the third order
polynomial equation (4) are of same sign 
within this parameter range.
%
%
%
\begin{figure}
\includegraphics[width=6.5cm]{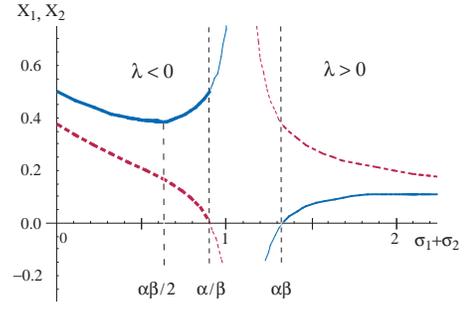}
\caption
{\label{fig2}
Fixed point coordinates
$X_1^\star$ and $X_2^\star$ as functions of $\sigma_1+\sigma_2$.
The parameters are $a_1 = 1$, $b_1=1$, $a_2=1.2$, $b_2=0.9$ and $d=2$.
Solid line represents $X_1^\star$ and the dahed $X_2^\star$.  
The fixed point is stable in the region
below $\alpha/\beta$ but unstable
above $\alpha \beta$.  The region
in between is unphysical.
 }
\end{figure}
%
%
%
\begin{figure}
\includegraphics[width=5.0cm]{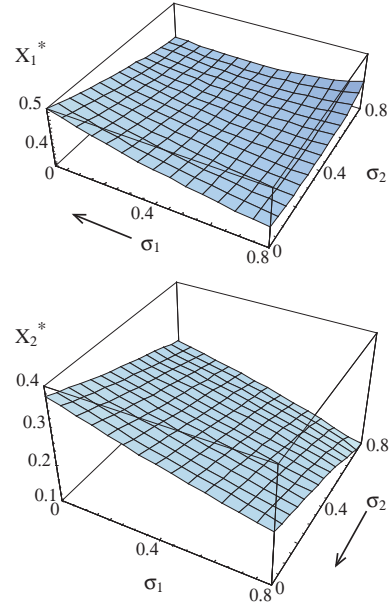}
\caption
{\label{fig3}
Fixed point coordinates
$X_1^\star$ and $X_2^\star$ as functions 
of $\sigma_1$ and $\sigma_2$.
The parameters are $a_1 = 1$, $b_1=1$, $a_2=1.2$, $b_2=0.9$ and $d=2$.
In the region $\sigma_1+\sigma_2$ $<\sigma_{cr}$ $=\alpha\beta/2$
$=a_2 b_1/b_2$, both $\sigma_1$ and $\sigma_2$ have to be decreased 
to make $X_1$ and $X_2$ larger.  
}
\end{figure}
%
%
%
%
\begin{figure}
\includegraphics[width=7.5cm]{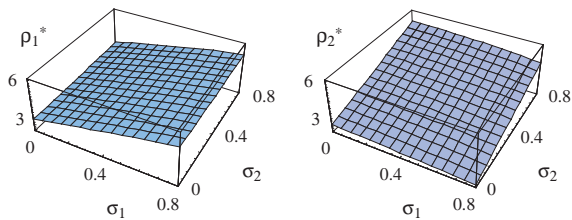}
\caption
{\label{fig4}
Aggression intensities
$\rho_1^\star$ and $\rho_2^\star$ as functions of $\sigma_1$ and $\sigma_2$.
The parameters are $a_1 = 1$, $b_1=1$, $a_2=1.2$, $b_2=0.9$, $d=2$
and $f=0.7$.
}
\end{figure}

The evolution of $\sigma_1$ and $\sigma_2$ depends on their
starting values.
With a straightforward calculation, we obtain
%
\begin{eqnarray}
{{\partial X_1^\star} \over {\partial \sigma_1}} < 0, \  
{{\partial X_2^\star} \over {\partial \sigma_2}} < 0
\ \ {\rm :} \ \ 
\sigma_1+\sigma_2 <  \sigma_{cr} ,
\\ \nonumber
{{\partial X_1^\star} \over {\partial \sigma_1}}  \cdot   
{{\partial X_2^\star} \over {\partial \sigma_2}} < 0
\ \ {\rm :} \ \ 
\sigma_1+\sigma_2 > \sigma_{cr} ,
\end{eqnarray}
within the range of (7).
The critical aggression intensity $\sigma_{cr}$ is given by
%
\begin{eqnarray}
\sigma_{cr}\equiv \max{\{\alpha\beta/2, \alpha/2 \beta\}}.
\end{eqnarray}
If the sum $\sigma_1+\sigma_2$ is bellow $\sigma_{cr}$, 
both $\sigma_1$ and $\sigma_2$ will move 
toward  $\sigma_1 =$ $\sigma_2 = 0$.
Namely, two competing species shall settle for a peaceful coexistence
as the common preys of a predator $y$. 
On the other hand, if the sum starts above critical value, 
$\sigma_1$ and $\sigma_2$ will
increase until one of the competing species is extinct at that critical value.
The situation becomes immediately clear 
with a glance at numerical
example depicted in Fig.\ \ref{fig2} and Fig.\ \ref{fig3}.

A crucial point is that the master $y$ acts as a punisher, according to (5),
that inhibits the increase of $\sigma_1$ and $\sigma_2$.
Fig.\ \ref{fig4} serves as a graphical illustration of this effect;
Increasing $\sigma_i$ will induce an increase of $\rho_i$ that 
incur the damage upon $x_i$.
We stress that no special mechanism is assumed
for $y$ to police the system in the outset, 
yet the dynamics seems to explain our common sense observation
about apex predators.

An intriguing fact is that the critical value $\sigma_{cr}$ is 
inversely proportional to the natural population 
of one of the prey species, $b_1/a_1$ or $b_2/a_2$.
This means that the coexistence of competing species
under common predator becomes a less likelier outcome
for a system with richer resources.  This seems to give
 a partial explanation to the experimentally observed
 decrease of species at the base levels
of trophic pyramids \cite{LB01}.

\begin{figure}
\includegraphics[width=2.5cm]{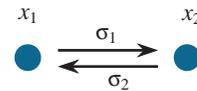}
\caption
{\label{fig5}
A symbolic diagram showing the two competing species
described by (1) with $\rho_1 = \rho_2 = 0$. 
}
\end{figure}

\begin{figure}
\includegraphics[width=5.0cm]{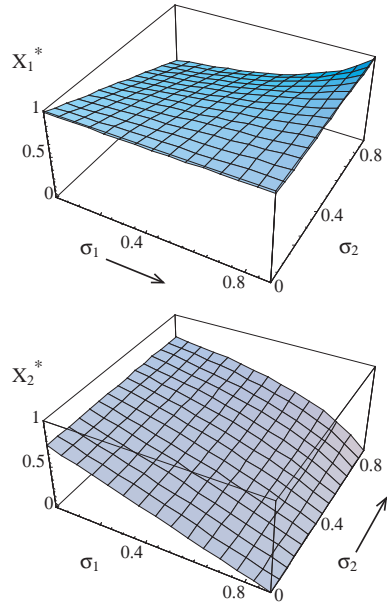}
\caption
{\label{fig6}
Fixed point coordinates
$X_1^\star$ and $X_2^\star$ for the case without 
the apex predator as functions of $\sigma_1$ and $\sigma_2$.
$X_1^\star$ will increase $\sigma_1$ and $X_2^\star$ will increase $\sigma_2$ to achieve
local advantage, which will eventually doom $X_2^\star$ to extinction.  
The parameters are $a_1 = 1$, $b_1=1$, $a_2=1.2$, and $b_2=0.9$.
}
\end{figure}

%
%
%
We next consider the case where the predator leaves the scene, namely $y = 0$
(Fig.\ \ref{fig5}).
By setting $\rho_1$ $=\rho_2$ $= 0$, we obtain, in place of (6), 
\begin{eqnarray}
X_1^\star &=&   { {a_2 b_1-\sigma_2 b_2} \over { a_1 a_2 - \sigma_1 \sigma_2 } }
\\ \nonumber
X_2^\star &=&   { {a_1 b_2-\sigma_1 b_1} \over { a_1 a_2 - \sigma_1 \sigma_2 } }.
\end{eqnarray}
The linearized map now takes a two-by-two matrix form
%
\begin{eqnarray}
M =
\begin{pmatrix}
-a_1 X_1   & -\sigma_2 X_1   \\
-\sigma_1 X_2   & -a_2 X_2  
\end{pmatrix} ,
\end{eqnarray}
in place of (3).
The straightforward calculation gives the condition for
$X_1^\star $ and $X_2^\star $ to be a viable fixed point, 
namely, $X_1^\star $, $X_2^\star  >0$, $\Re\lambda < 0$, in terms
of the allowed region for the aggression intensity as
%
\begin{eqnarray}
\sigma_1 < \min{\{\alpha, \alpha\beta\}}, \ \ 
\sigma_2 < \min{\{\alpha, \alpha/\beta\}} .
\end{eqnarray}
However, within this region, we can easily check the relation
\begin{eqnarray}
{{\partial X_1^\star } \over {\partial \sigma_1}} > 0, \ \
{{\partial X_2^\star } \over {\partial \sigma_2}} > 0.
\end{eqnarray}
Therefore, in this case, both $\sigma_1$ or $\sigma_2$ shall eventually be increased 
beyond the range (12), 
and there is no evolutionarily stable coexisting solutions for two
competing species.
Namely, in the absence of the common master, one of the competing species
is always driven to extinction by arms race of
increasing $\sigma_1$ and $\sigma_2$.  
An example of this case is illustrated in  Fig.\ \ref{fig6}
%
%
%
%
%
%

%

In order to fully understand the generic shape of the ecosystem,
we would have to generalize  our arguments
to more trophic levels than two, and also more species than two
within a single trophic level. 
To this end, this work, in combination to our previous work, is but a
modest start.
More involved and sophisticated approaches
of both numerical and analytical nature 
may have to be called for \cite{LB01, OF02, CS03}.
In the current work, no precise specification for the
evolutionary dynamics of behavioral parameters has been required.
While we stress that this is an advantage,
construction and analysis of more detailed 
models with such specifications are attractive possibilities.  

Finally, some remarks in the broader context of game theory  \cite{MS82}
are in order.
There is an obvious game-theoretic interpretation
of the results (6) and (10).
For the sake of simplicity, let us
 set $a_1 = a_2 = a$ and $b_1 = b_2 = b$.
We now regard $\sigma_i$ ($i = 1, 2$) as the control parameter of
the strategy of population $x_i$ for the game played between $x_1$
and $x_2$ whose payoff tables are given 
by $X_1^\star[\sigma_1,\sigma_2]$ and $X_2^\star[\sigma_1,\sigma_2] $.
To facilitate the understanding, the game tables
for two discretized points for $\sigma_1$ and $\sigma_2$ are tabulated 
in Table I.
For the case without the common predator, (10),
the game is a continuous strategy version of 
prisoner's dilemma \cite{AX84, GI00},
whose outcome is $\sigma_1 \to a$, $\sigma_2 \to a$ which leads
to the extinction of either $x_1$ or $x_2$.  
With the introduction of the apex predator, the game table is turned into
one of collaborative game, whose outcome is the coexistence
$\sigma_1 \to 0$ and  $\sigma_2 \to 0$.
Note that the game table is symmetrized under the presence of the apex 
predator;
$X_2^\star[\sigma_1,\sigma_2]$ = $X_1^\star[\sigma_2,\sigma_1]$.
This could be interpreted as the sign of altruistic behavior \cite{CH03a}.
Advantage of having the apex predator as a ``law enforcer'' 
is evident, and the loss of half of the populations to the predation 
would be an acceptable  tradeoff. 
%
%
%
\begin{table}
\label{tab1 }
\caption{
The game table $X_1^\star[\sigma_1,\sigma_2]$ for $x_1$ discretized 
at $\sigma_1$, $\sigma_2$ $=$ $a/3$
(low aggressiveness, ``dove'')
and $2a/3$ (high aggressiveness, ``hawk'').  
Left hand side is the table for the case with apex predator (6), and
the right hand side, without apex predator (10).  
The game table $X_2^\star[\sigma_1,\sigma_2]$ for $x_2$ is obtained
by transposition of raw and column.   
The Nash equilibrium is indicated with boldface. }
{\ \  }\\
{With Apex Predator \ \ \
\qquad
No Apex Predator}\\
{\ \  }\\
\begin{tabular}{c|cc}
  { ${}_{\sigma_1} \backslash {}^{\sigma_2}$ }   & 
  { ${\ }^{a/3}_{dove}$ }  &
  { ${\ }^{2a/3}_{hawk}$ }   \\
  & & \\
\hline
  & & \\
  { ${}^{ a/3}_{dove}$ }  & 
  { ${\bf { 3b \over {8a}} }$ } &
  {  ${ { 3b} \over {9a} } $  }\\
  & & \\
 { ${}^{2a/3}_{hawk}$} & 
 { ${{3b}\over{9a}}$ } &
 { ${ { 3b} \over {10a} }$ } \\
\end{tabular}
\quad
%
%
\begin{tabular}{c|cc}
  { ${}_{\sigma_1} \backslash {}^{\sigma_2}$ }   & 
  { ${\ }^{a/3}_{dove}$ }  &
  { ${\ }^{2a/3}_{hawk}$ }   \\
  & & \\
\hline
  & & \\
 { ${}^{ a/3}_{dove}$}  & 
 { ${ 6b \over {8a}}$ } &
 { ${ { 6b} \over {14a} }$ } \\
  & & \\
 { ${}^{2a/3}_{hawk}$} & 
 { ${{6b}\over{7a}}$ } &
 { $ {\bf { { 6b} \over {10a} }  }$ }\\
\end{tabular}
\end{table}
%
%
%
%
%
%

In summary, we have established, for Lotka-Volterra
systems with evolutionary parameter variation,  
that two competing species are evolutionarily unstable, but
can be stabilized by the introduction of an apex predator. 
We hope this to be a start for systematic understanding 
of stable ecosystems.

%
The authors wish to express gratitudes to 
Professors
Kazuo Takayanagi,
Toshiya Kawai 
and David Greene 
for helpful discussions and useful comments.

%
%

\end{document}